\documentclass{ws-ijmpa}
\usepackage{%enumerate,svsing2e,
amsmath%,amsfonts,latexsym,graphicx,
}

\begin{document}
%\markboth{Authors' Names}
%\catchline{}{}{}

\title{LIGHT PROPAGATION IN\\ GENERALLY COVARIANT ELECTRODYNAMICS\\ 
  AND THE FRESNEL EQUATION\footnote{Invited talk given at Journe\'es
    Relativistes, University College Dublin, Sept.\ 2001. Dedicated to
    the memory of Rev.\ Dr.\ J.\ Dermott McCrea, OFM, who passed away
    on 21 May 1993.}}

\author{Friedrich W.\ Hehl\footnote{Email: hehl@thp.uni-koeln.de.} ,
  Yuri N.\ Obukhov\footnote{Email: yo@thp.uni-koeln.de. Permanent
    address: Department of Theoretical Physics, Moscow State
    University, 117234 Moscow, Russia.} , and Guillermo F.\ 
  Rubilar\footnote{Email: gr@thp.uni-koeln.de.}} \address{Institute
  for Theoretical Physics, University of Cologne\\ 50923 K\"oln,
  Germany}

\maketitle

%\pub{Received 04 March 2002}{Communicated by P.A.\ Hogan}
%Revised (Day Month Year)}

\begin{abstract}
  Within the framework of generally covariant (pre-metric)
  electrodynamics, we specify a local vacuum spacetime relation
  between the excitation $H=({\cal D},{\cal H})$ and the field
  strength $F=(E,B)$. We study the propagation of electromagnetic
  waves in such a spacetime by Hadamard's method and arrive, with the
  constitutive tensor density $\kappa\sim\partial H/\partial F$, at a
  Fresnel equation which is algebraic of 4th order in the wave
  covector. We determine how the different pieces of $\kappa$, in
  particular the axion and the skewon pieces, affect the propagation
  of light.\hfill{\it file dublinws6.tex, 2002-03-27}
\end{abstract}

\section{Introduction}

Electromagnetic wave propagation is a very important physical
phenomenon in classical field theory. In general, the geometrical
structure of spacetime as well as the intrinsic properties and the
motion of material media can affect the light propagation. In the
generally covariant pre-metric approach to
electrodynamics\cite{Schouten,Post,HO02}, the axioms of electric
charge and of magnetic flux conservation manifest themselves in the
Maxwell equations for the excitation $H=({\cal D},{\cal H})$ and the
field strength $F=(E,B)$:
\begin{equation}\label{me}
dH=J, \qquad dF=0 \,.
\end{equation}
These equations should be supplemented by a constitutive law $H =
H(F)$.  The latter relation contains the crucial information about the
underlying physical continuum (i.e., spacetime and/or material
medium). Mathematically, this constitutive law arises either from a
suitable phenomenological theory of a medium or from the
electromagnetic field Lagrangian. It can be a nonlinear or even
nonlocal relation between the electromagnetic excitation and the field
strength.

Earlier, we have investigated the propagation of waves in the most
general linear theory. Here we present some results which hold true
for all electrodynamical models with an arbitrary local spacetime
relation.  This is how we call the constitutive law if it applies to
spacetime (``the vacuum'') itself.

If local coordinates $x^i$ are given, with $i,j,... =0,1,2,3$, we can
decompose the excitation and field strength 2-forms into their
components according to
\begin{equation}
H = {\frac 1 2}\,H_{ij}\,dx^i\wedge dx^j,\qquad
F = {\frac 1 2}\,F_{ij}\,dx^i\wedge dx^j.\label{geo1}
\end{equation}

\section{Wave propagation: Fresnel equation}

We will study the propagation of a discontinuity of the
electromagnetic field following the lines of Ref.\ \!\cite{OFR00}, see
also Refs.\ \!\cite{nonsym32,Dr.Guillermo}.  The surface of
discontinuity $S$ is defined locally by a function $\Phi$ such that
$\Phi= const$ on $S$. Across $S$, the geometric Hadamard conditions
are satisfied:
\begin{eqnarray}
&& [F_{ij}] = 0,\qquad [\partial_i F_{jk}] = q_i\, f_{jk}, 
\label{had1}\\
&& [H_{ij}] = 0,\qquad [\partial_i H_{jk}] = q_i\, h_{jk}. 
\label{had2}
\end{eqnarray}
Here $\left[{\cal F}\right](x)$ denotes the discontinuity of a
function ${\cal F}$ across $S$, and $q_i:=\partial_i\Phi$ is the wave
covector.  Since the spacetime relation $H(F)$ determines the
excitation in terms of the field strength, the corresponding tensors
$f_{ij}$ and $h_{ij}$, describing the jumps of the derivatives of
field strength and excitation, are related by
\begin{equation}\label{kappa}
h_{ij} = {\frac 1 2}\,\kappa_{ij}{}^{kl}\,f_{kl},\qquad {\rm with}
\qquad \kappa_{ij}{}^{kl} := {\frac {\partial H_{ij}}{\partial F_{kl}}}.
\end{equation}
In linear electrodynamics, the components of the constitutive tensor
$\kappa_{ij}{}^{kl}$ are constant (or, at least, independent of the
electromagnetic field). But in general $\kappa_{ij}{}^{kl}$ is a
function of the electromagnetic field, the velocity of matter, the
temperature, and other physical and geometrical variables. Quite
remarkably, however, all the earlier results obtained for linear
electrodynamics remain also valid in the general case because whatever
nonlinear spacetime relation $H(F)$ may exist, the relation between
the {\it jumps} of the field derivatives, according to (\ref{kappa}),
is always linear.

If we use Maxwell's equations (\ref{me}), then (\ref{had1}) and 
(\ref{had2}) yield
\begin{equation}\label{4Dwave}
  {\epsilon}^{\,ijkl}\, q_{j}\,h_{kl}=0 \,,\qquad
  {\epsilon}^{\,ijkl}\, q_{j}\,f_{kl}=0\,.
\end{equation}
Introducing the conventional constitutive matrix
\begin{equation}
\chi^{ijkl} = {\frac 1 2}\,\epsilon^{ijmn}\,\kappa_{mn}{}^{kl}
\end{equation}
and making use of (\ref{kappa}), we rewrite the above system as
\begin{equation}\label{4Dwave1}
  {\chi}^{\,ijkl}\, q_{j}\,f_{kl}=0 \,,\qquad
  {\epsilon}^{\,ijkl}\, q_{j}\,f_{kl}=0\,.
\end{equation}
Solving the last equation by $f_{ij} = q_ia_j - q_ja_i$, we finally
reduce (\ref{4Dwave1})$_1$ to
\begin{equation}\label{4Dwave3}
  {\chi}{}^{\,ijkl}\,q_{j}q_ka_l=0 \,.
\end{equation}
This algebraic system has a nontrivial solution for $a_i$ only when 
the determinant of the matrix on the l.h.s. vanishes. The latter
gives rise to our {\it covariant Fresnel equation}
\begin{equation} \label{Fresnel}  
{\cal G}^{ijkl}(\chi)\,q_i q_j q_k q_l = 0 \,,
\end{equation}
with the fourth order Fresnel tensor density of weight $+1$
defined by
\begin{equation}\label{G4}  
  {\cal G}^{ijkl}(\chi):=\frac{1}{4!}\,{\epsilon}_{mnpq}\,
  {\epsilon}_{rstu}\, {\chi}^{mnr(i}\, {\chi}^{j|ps|k}\,
  {\chi}^{l)qtu }\,.
\end{equation}
It is totally symmetric, ${\cal G}^{ijkl}(\chi)= {\cal
  G}^{(ijkl)}(\chi)$, and thus has  35 independent components.

\section{Spacetime relation: The emergence of the axion and the 
skewon fields}

The quantity $\kappa_{ij}{}^{kl}$ or, equivalently, $\chi^{ijkl}(x)$,
characterizes the electromagnetic properties of the vacuum and is as
such of universal importance. The untwisted tensor density
$\chi^{ijkl}(x)$ of weight $+1$ has 36 independent components.  If we
decompose it into irreducible pieces with respect to the 6-dimensional
linear group, then we find
\begin{equation}\label{decomp'}\chi^{ijkl}={}^{(1)}\chi^{ijkl}
  + {}^{(2)}\chi^{ijkl} + {}^{(3)}\chi^{ijkl}\,, \quad{\rm with} \quad
  36= 20\oplus 15 \oplus 1\end{equation} independent components,
respectively. The irreducible pieces of $\chi$ are defined as follows:
\begin{eqnarray}\label{chiirr2}
  {}^{(2)}\chi^{ijkl}&:=&\frac{1}{2}\left( \chi^{ijkl}-
    \chi^{klij}\right)=-{}^{(2)}\chi^{klij}\,,\quad
  {}^{(3)}\chi^{ijkl}:=\chi^{[ijkl]}\,, \nonumber\\{}^{(1)}
  \chi^{ijkl}&:=& \chi^{ijkl} -\,
  {}^{(2)}\chi^{ijkl}-\,{}^{(3)}\chi^{ijkl}={}^{(1)} \chi^{klij}\,.
\end{eqnarray}
Since no metric is available, we cannot form traces.  The Abelian
axion piece $^{(3)}\chi^{ijkl}$ $=:\alpha(x)\,\epsilon^{ijkl}$ has
been introduced by Ni \cite{Ni73,Ni77}. A constitutive law (for
matter) with $^{(2)}\chi^{ijkl}\ne0$ (non-vanishing ``skewon field'')
has been discussed by Nieves and Pal\cite{NP89,NP94}. It yields P-
and CP-violating terms in the field equations. Thus all constitutive
functions in (\ref{decomp'}) can claim respectability from a physical
point of view in the framework of linear response theory.

Alternatively, by means of mere contractions, we can decompose
\begin{eqnarray}
\kappa_{ij}{}^{kl} &=& {}^{(1)}\kappa_{ij}{}^{kl} +
  {}^{(2)}\kappa_{ij}{}^{kl} + {}^{(3)}\kappa_{ij}{}^{kl}\\ &=&
  {}^{(1)}\kappa_{ij}{}^{kl} +
  2\!\not\!\kappa_{[i}{}^{[k}\,\delta_{j]}^{l]} + {\frac 1
    6}\,\kappa\,\delta_{[i}^k\delta_{j]}^l\,.\label{kap-dec}
\end{eqnarray}
Here $\kappa_i{}^k := \kappa_{il}{}^{kl}$, $\kappa := \kappa_k{}^k = 
\kappa_{kl}{}^{kl}$, and the traceless piece is $\not\!\kappa_i{}^k := 
\kappa_i{}^k - {\frac 1 4}\,\kappa\,\delta_i^k$. 

By introducing the pseudotensorial skewon field and the pseudoscalar
axion field by
\begin{equation}
  S_i{}^j = -\,{\frac 1 2}\!\not\!\kappa_i{}^j=\not\!\! S_i{}^j,\qquad
  \alpha = {\frac 1 {12}}\,\kappa,
\end{equation}
we ultimately find the basic relations between the corresponding
irreducible pieces:
\begin{eqnarray}
{}^{(1)}\chi^{ijkl} &=& {\frac 1 2}\,\epsilon^{ijmn}\,\,{}^{(1)}
\kappa_{mn}{}^{kl},\\ 
{}^{(2)}\chi^{ijkl} &=& {\frac 1 2}\,\epsilon^{ijmn}\,\,{}^{(2)}
\kappa_{mn}{}^{kl} = \epsilon^{ijm[k}\,S_m{}^{l]} -
\epsilon^{klm[i}\,S_m{}^{j]},\label{skewon}\\
{}^{(3)}\chi^{ijkl} &=& {\frac 1 2}\,\epsilon^{ijmn}\,\,{}^{(3)}
\kappa_{mn}{}^{kl} = \alpha\,\epsilon^{ijkl}\,. 
\end{eqnarray}

\section{Properties of the Fresnel tensor density}

The various irreducible pieces of the constitutive tensor affect the
wave propagation in different ways. Technically, this can be
determined by studying how a certain irreducible piece ${}^{(a)}\chi$
contributes to the Fresnel tensor density (\ref{G4}) and, thereby, to
the Fresnel equation (\ref{Fresnel}) which governs the wave covectors.

We will demonstrate here the general properties of the Fresnel tensor
density. As a preparatory step, let us take $\chi = \phi + \psi$.
Then, using a compact notation by omitting the indices, we have quite
generally
\begin{equation} \label{over1}
{\cal G}(\chi) = {\cal G}(\phi) + {\cal G}(\psi) 
+ {\frac 1{4!}}\left(O_1 + O_2 + O_3 + T_1 + T_2 + T_3\right).
\end{equation}
Here the mixed terms $O_a$ contain one $\psi$-factor and the $T_a$'s
two $\psi$-factors. {\it Postponing the symmetrization} over $i,j,k,l$
to the very last moment, these terms read explicitly as follows:
\begin{eqnarray}
O_1(\phi,\psi,\phi) &=& {\epsilon}_{mnpq}\,{\epsilon}_{rstu}\, 
\phi^{mnri}\,\psi^{jpsk}\,\phi^{lqtu}\,,\label{o1}\\
O_2(\psi,\phi,\phi) &=& {\epsilon}_{mnpq}\,{\epsilon}_{rstu}\, 
\psi^{mnri}\,\phi^{jpsk}\,\phi^{lqtu}\,,\label{o2}\\
O_3(\phi,\phi,\psi) &=& {\epsilon}_{mnpq}\,{\epsilon}_{rstu}\,
\phi^{mnri}\,\phi^{jpsk}\,\psi^{lqtu}\,,\label{o3}\\ 
T_1(\psi,\phi,\psi) &=& {\epsilon}_{mnpq}\,{\epsilon}_{rstu}\, 
\psi^{mnri}\,\phi^{jpsk}\,\psi^{lqtu}\,,\label{t1}\\
T_2(\psi,\psi,\phi) &=& {\epsilon}_{mnpq}\,{\epsilon}_{rstu}\, 
\psi^{mnri}\,\psi^{jpsk}\,\phi^{lqtu}\,,\label{t2}\\
T_3(\phi,\psi,\psi) &=& {\epsilon}_{mnpq}\,{\epsilon}_{rstu}\,
\phi^{mnri}\,\psi^{jpsk}\,\psi^{lqtu}\,.\label{t3}
\end{eqnarray}

Now we are in position to prove the following properties of the Fresnel
tensor:
\medskip

\noindent{\bf 1)} ${\cal G}({}^{(3)}\chi) =0$. 
{\it Proof\/}: By direct substitution of ${}^{(3)}\chi^{ijkl} = \alpha
\,\epsilon^{ijkl}$, we have
\begin{eqnarray}
{\epsilon}_{mnpq}\,{\epsilon}_{rstu}\,{}^{(3)}\chi^{mnri}\,{}^{(3)}
\chi^{jpsk}\,{}^{(3)}\chi^{lqtu} &=& 4\alpha^3\,(\delta_p^r\delta_q^i - 
\delta_p^i\delta_q^r)(\delta_r^l\delta_s^q - \delta_r^q\delta_s^l)\,
\epsilon^{jpsk}\nonumber\\
&=& 12\alpha^3\,\epsilon^{jilk}.\nonumber
\end{eqnarray}
This vanishes upon symmetrization over the indices $i,j,k,l$.
\medskip

\noindent{\bf 2)} ${\cal G}({}^{(2)}\chi) =0$.
{\it Proof\/}: Using the symmetry properties (\ref{chiirr2}) and
especially the skew symmetry $^{(2)}{\chi}^{ijkl} =
-\,^{(2)}{\chi}^{klij}$, we find
\begin{eqnarray}
{\epsilon}_{mnpq}\,{\epsilon}_{rstu}\,{}^{(3)}\chi^{mnri}\,{}^{(3)}
\chi^{jpsk}\,{}^{(3)}\chi^{lqtu} \!&=&\! -\,{\epsilon}_{mnpq}
\,{\epsilon}_{rstu}\,{}^{(3)}\chi^{rimn}\,{}^{(3)}\chi^{skjp}
\,{}^{(3)}\chi^{tulq}\nonumber\\
\!&=&\! -\,{\epsilon}_{mnpq}\,{\epsilon}_{rstu}\,{}^{(3)}\chi^{tuql}
\,{}^{(3)}\chi^{kspj}\,{}^{(3)}\chi^{irmn}\nonumber\\
\!&=&\! -\,{\epsilon}_{mnpq}\,{\epsilon}_{rstu}\,{}^{(3)}\chi^{mnrl}
\,{}^{(3)}\chi^{kpsj}\,{}^{(3)}\chi^{iqtu}.\nonumber
\end{eqnarray}
Upon the symmetrization over the indices $i,j,k,l$, we then get ${\cal
  G}({}^{(2)}\chi) = -\,{\cal G}({}^{(2)}\chi)$ or ${\cal
  G}({}^{(2)}\chi) =0$.  \medskip

\noindent{\bf 3)} ${\cal G}({}^{(1)}\chi + {}^{(2)}\chi + {}^{(3)}\chi) = 
{\cal G}({}^{(1)}\chi + {}^{(2)}\chi)$. {\it Proof\/}: Let us put
$\phi = {}^{(1)}\chi + {}^{(2)}\chi$ and $\psi = {}^{(3)}\chi$ in the
formula (\ref{over1}). Then we can verify that all the mixed terms
(\ref{o1})-(\ref{t3}) vanish. Indeed, using ${}^{(3)}\chi^{ijkl} =
\alpha\,\epsilon^{ijkl}$, we find:
\begin{eqnarray}
O_1 &=& 2\alpha\,{\epsilon}_{mnpq}\,\phi^{mnri}\,\phi^{lqtu}
\,(\delta_r^j\delta_t^p\delta_u^k + \delta_r^p\delta_t^k\delta_u^j 
+ \delta_r^k\delta_t^j\delta_u^p)\nonumber\\
&=& 2\alpha\,{\epsilon}_{mnpq}\,(\phi^{mnji}\,\phi^{lqpk}
+ \phi^{mnpi}\,\phi^{lqkj} + \phi^{mnki}\,\phi^{lqjp}).\nonumber
\end{eqnarray}
This is zero when we impose the symmetrization over $i,j,k,l$.
Similarly, we find
\begin{equation}
O_2 = -\,2\alpha\,{\epsilon}_{rstu}\,(\delta_p^r\delta_q^i - 
\delta_p^i\delta_q^r)\phi^{jpsk}\,\phi^{lqtu} 
= 2\alpha\,{\epsilon}_{rstu}\,(-\,\phi^{jrsk}\,\phi^{litu} 
+ \phi^{jisk}\,\phi^{lrtu})\nonumber
\end{equation}
and 
\begin{equation}
O_3 = -\,2\alpha\,{\epsilon}_{mnpq}\,(\delta_r^l\delta_s^q - 
\delta_r^q\delta_s^l)\phi^{mnri}\,\phi^{jpsk} 
= 2\alpha\,{\epsilon}_{mnpq}\,(-\,\phi^{mnli}\,\phi^{jpqk} 
+ \phi^{mnqi}\,\phi^{jplk}).\nonumber
\end{equation}
Both expressions vanish when we impose symmetrization over
$i,j,k,l$.  The proof that all $T$'s are equal to zero reduces to the
above formulas in which we simply need to replace one of the $\phi$
factors by the Levi-Civita $\epsilon$. Consequently, (\ref{over1})
yields ${\cal G}(\phi + \psi) = {\cal G}(\phi) + {\cal G}(\psi) =
{\cal G}(\phi)$, since ${\cal G}(\psi) = {\cal G}(^{(3)}{\chi}) =0$.
\medskip

\noindent {\bf 4)} ${\cal G}({}^{(2)}\chi + {}^{(3)}\chi) = 0$.
{\it Proof\/}: Take $\phi = {}^{(2)}\chi$ and $\psi = {}^{(3)}\chi$.
Then the proof reduces to the above case. [Formally, we can simply
consider a special case of the formula ${\cal G}({}^{(1)}\chi +
{}^{(2)}\chi + {}^{(3)}\chi) = {\cal G}({}^{(1)}\chi + {}^{(2)}\chi)$
by putting ${}^{(1)}\chi =0$ and using the property ${\cal
  G}(^{(2)}{\chi}) =0$].

\section{Discussion and conclusion}

The properties of the Fresnel tensor simplify greatly the final analysis 
of the light propagation in the general electrodynamical theory. 

As we saw, neither the axion piece $^{(2)}{\chi}$ nor the skewon piece
$^{(3)}{\chi}$ (alone or together) can provide electromagnetic wave
propagation.

The presence of the {\it principal part} $^{(1)}{\chi}$ of the
constitutive tensor is indispensable for the existence of nontrivial
electromagnetic waves and ultimately for the existence of the light
cone structure on spacetime. In any case, the Fresnel tensor density
reads
\begin{equation}\label{propg1}
 {\cal G}^{ijkl}(\chi)= {\cal G}^{ijkl}({}^{(1)}\chi+{}^{(2)}\chi) .
\end{equation}
Furthermore, in general
\begin{equation} \label{propg5}
  {\cal G}^{ijkl}({}^{(1)}\chi+ {}^{(2)}\chi)\neq {\cal
    G}^{ijkl}({}^{(1)}\chi) ,
\end{equation}
which means that the skewon field {\it does} influence the Fresnel
equation and thus, eventually, the light cone structure. An example of
this general result can be found in the asymmetric constitutive tensor
studied by Nieves and Pal \cite{NP89,NP94}.
  
Actually, we can use (\ref{over1}) to find more exactly the contribution
of the skewon to the Fresnel tensor. One can straightforwardly see that
\begin{eqnarray}
O_1({}^{(1)}\chi, {}^{(2)}\chi, {}^{(1)}\chi) &=& 0,\quad 
O_2({}^{(2)}\chi, {}^{(1)}\chi, {}^{(1)}\chi) + 
O_3({}^{(1)}\chi, {}^{(1)}\chi, {}^{(2)}\chi) =0,\\
T_2({}^{(2)}\chi, {}^{(2)}\chi, {}^{(1)}\chi) &=& 
T_3({}^{(1)}\chi, {}^{(2)}\chi, {}^{(2)}\chi).
\end{eqnarray}
The proof is analogous to the demonstration of the property ${\cal
  G}({}^{(2)}\chi) =0$ and is based directly on the skew symmetry
$^{(2)}{\chi}^{ijkl} = -\,^{(2)}{\chi}^{klij}$. As a result,
(\ref{over1}) yields
\begin{equation}
{\cal G}({}^{(1)}\chi+ {}^{(2)}\chi) = 
{\cal G}({}^{(1)}\chi) + {\frac 1 {4!}}\,[
T_1({}^{(2)}\chi, {}^{(1)}\chi, {}^{(2)}\chi) + 
2\,T_2({}^{(2)}\chi, {}^{(2)}\chi, {}^{(1)}\chi)].\nonumber
\end{equation}
Substituting here the explicit form of the skewon part (\ref{skewon}),
we finally find
\begin{equation} \label{propg8}
  {\cal G}^{ijkl}(\chi) = {\cal G}^{ijkl}({}^{(1)}\chi + {}^{(2)}\chi)
  = {\cal G}^{ijkl}({}^{(1)}\chi) + {}^{(1)}\chi^{\,m(i|n|j}S_m^{\ k}
  S_n^{\ l)}\,.
\end{equation}

Our basic results on the light propagation are then that
$^{(1)}\chi\neq 0$, otherwise there is no orderly wave propagation,
the axion piece $\alpha$ is left arbitrary, it doesn't influence light
propagation locally, whereas the skewon piece $S_i{}^j=\not\!\! S_i{}^j$
makes itself felt by means of equation (\ref{propg8}). The detailed
study of how exactly these 15 functions $\not\!\! S_i{}^j$ affect the light
cone is left for future work.

\section*{Acknowledgments}
We are grateful to Peter Hogan for the invitation to give a talk at
Journe\'es Relativistes 2001. We also thank Peter for his splendid
hospitality. G.F.R.\ would like to thank the German Academic Exchange
Service (DAAD) for financial support.

%\appendix

%\section{Appendices}

\end{document}